# Spin Hall magnetoresistance in Pt/YIG bilayers via varying magnon excitation

Q. B. Liu[1,2], K. K. Meng[1]*, S. Q. Zheng[1], Y. C. Wu[1], J. Miao[1], X. G. Xu[1] and Y. Jiang[1]**

[1]*Beijing Advanced Innovation Center for Materials Genome Engineering, School of Materials Science and Engineering, University of Science and Technology Beijing, Beijing 100083, China*

[2]*Applied and Engineering Physics, Cornell University, Ithaca, NY 14853, USA*

**Abstract:** Spin Hall magnetoresistance (SMR) and magnon excitation magnetoresistance (MMR) that all generate via the spin Hall effect and inverse spin Hall effect in a nonmagnetic material are always related to each other. However, the influence of magnon excitation for SMR is often overlooked due to the negligible MMR. Here, we investigate the SMR in Pt/$Y_3Fe_5O_{12}$ (YIG) bilayers from 5 to 300K, in which the YIG are treated after $Ar^+$-ion milling. The SMR in the treated device is smaller than in the non-treated. According to theoretical simulation, we attribute this phenomenon to the reduction of the interfacial spin-mixing conductance at the treated Pt/YIG interface induced by the magnon suppression. Our experimental results point out that the SMR and the MMR are inter-connected, and the former could be modulated via magnon excitation. Our findings provide a new approach for separating and clarifying the underlying mechanisms.

*Authors to whom correspondence should be addressed:
*kkmeng@ustb.edu.cn

**yjiang@ustb.edu.cn

Ferromagnetic insulators (FMIs) has emerged as a promising and novel technology platform to generate, modulate and detect spin information over long distances [1,2]. The advantage of using FMIs against metallic ones is avoiding the flow of charge current, preventing ohmic losses and the emergence of undesired spurious effects [3]. Yttrium Iron Garnet (YIG) is one of the most prominent FMIs for the investigation of spintronics, magnonics [4,5], and spin caloritronics [6,7], due to its extremely low damping, soft ferrimagnetism and negligible in-plane magnetic anisotropy even in limit thickness. One can efficiently investigate and control the magnetization direction and spin waves propagation in FMIs through transport measurements based on the spin Hall effect (SHE) and the inverse spin Hall effect (ISHE) [2,8,9], by which the mutual conversion between magnon or spin current and charge current can be realized in a heavy metal (HM) with strong spin-orbit coupling (SOC). Therefore, the combination of SHE and ISHE can give rise to the spin Hall magnetoresistance (SMR) and magnon excitation magnetoresistance (MMR) [10,11], while the former can be ascribed to the spin transfer torque at the HM/FMIs interface. Owing to the SHE, the spin current would accumulate at the HM/FMIs interface with spin polarization $\sigma$ parallel to the surface. If $\sigma$ is not collinear with the magnetization $M$ of the FMIs, the accumulated spin current can exert a torque proportional to $M\times(M\times\sigma)$ on the $M$ of FMIs. It suggests that a finite spin current will be absorbed by FMIs if $M$ and $\sigma$ have a finite angle. The spin current absorption by FMIs represents the spin current reflection that is suppressed, and its resistance therefore will change with the magnetization orientation which is expected to be maximized (minimized) when $M$ is perpendicular (parallel) to $\sigma$ [8,9]. On the other hand, the MMR is firstly observed in Pt/YIG heterostructures in which the two parallel Pt strips were separated by a distance. When the two Pt electrodes are closed enough, the spin accumulation at one Pt strip will transmit into another strip by magnon, and the spin current will be converted into charge current via ISHE. Therefore, a large non-local charge current is supposed for $M//\sigma$, since the magnons beneath the first Pt strip can diffuse across the gap to the second Pt strip. In contrast, for $M\perp\sigma$, the non-local signals should be significantly reduced, since the spin transfer torque absorbs the magnon propagation. Therefore, the measured resistance or voltage signal is expected to be maximized (minimized) when $M$ is parallel (perpendicular) to $\sigma$ [12]. The MMR can thus be seen as a magnon-mediated, non-local counterpart of the SMR. In short, both the SMR and the MMR stem from spin accumulation and spin transport at the HM/FIMs interfaces. Theoretically, the efficiency and strength of the spin transfer across the interface depend on the magnitude of interfacial spin-mixing conductance (SMC) $G_{\uparrow\downarrow}=(G_r+iG_i)$, where $G_r$ ($G_i$) denotes the real (imaginary) part [13,14]. Therefore, if the magnon transport can be suppressed or enhanced, it will change the magnitude of $G_r$ and the SMR is expected to change correspondingly. However, the influence of magnon excitation for SMR is often overlooked due to the negligible MMR, and few methods have been reported to separate and clarify the underlying mechanisms.

In this work, we have employed the etching treatment to suppress the in-plane magnon transport and found that the SMR in Pt/YIG-based devices was dramatically altered when the part of YIG film out of the hall devices is treated with Ar+-ion milling etching. At room temperature, the SMR in the treated device was much smaller than that in the non-treated, while at low temperature the SMR for the two devices was similar. The anomalous SMR reduction has similar temperature dependence with magnon excitation. According to the theoretical simulation, we have attributed this phenomenon to the reduction of the $G_{\uparrow\downarrow}$ at the Pt/YIG interface induced by magnon suppression after etching. Our experimental results pointed out that the SMR and the MMR were interconnected, while the former can be modulated via magnon excitation even though the value due to magnon excitations was much smaller than the SMR.

The epitaxial YIG film was grown on a [111]-oriented GGG substrate (lattice parameter $a$ = 1.237 nm) by pulsed laser deposition technique with the substrate temperature $T_S$ = 780℃ and the oxygen pressure 10 Pa. Then the samples were annealed at 780℃ for 30 min at the oxygen pressure of 200 Pa. The base pressure of the PLD cavity was better than $2 \times 10^{-6}$ Pa. Then, the Pt layer was deposited on YIG at room temperature by magnetron sputtering. In order to avoid the run-to-run error, each large Pt/YIG sample was then cut into two small pieces. After the deposition, the electron beam lithography and Ar ion milling were used to pattern Hall bars, and a lift-off process was used to form contact electrodes. The size of all the Hall bars is 20 μm×120 μm. For comparison, a part of YIG film out of the Hall bars was etched away by $Ar^+$-ion milling which was defined as $YIG^+$ as shown in Fig. 1(a). Fig. 1 (b) shows the XRD *ω-2θ* scan spectra of the 40-nm-thick YIG thin film, which was taken from representative thin film of each type, and it shows predominant (444) diffraction peaks with no diffraction peaks occurring from impurity phases or other crystallographic orientation, indicating the single phase nature. According to the (444) diffraction peak position and the reciprocal space maps of the (642) reflection of 30-nm-thick YIG films grown on GGG as shown in Fig. 1(c), we have found that the lattice constant of YIG layer is similar with the value of GGG substrate, indicating the high quality epitaxial growth without mismatch. Moreover, the saturated magnetization of the YIG layer measured by a vibrating sample magnetometer was determined to be 140 emu/cm$^3$ as shown in Fig. 1(d), which is similar with the value of the bulk YIG. All the magnetotransport measurements performed in the multilayers were carried out using a Keithley 6221 sourcemeter and a Keithley 2182A nanovoltmeter. These measurements were performed at different temperatures from 5 and 300 K in a liquid-He cryostat that allows applying magnetic fields $H$ up to 3 T and rotating the samples by $360°$.

Using small and non-perturbative current densities (~ $10^6$ A/cm$^2$), we have investigated the angular-dependent magnetoresistance (ADMR) in Pt (5 nm)/YIG (40 nm) and Pt (5 nm)/YIG$^+$ (40 nm) devices at room temperature. The measurement configuration, the definition of the axes, and the rotation angles ($α$, $β$, $γ$) are defined in the sketches as shown in Fig. 2(a). Figs. 2(b) and (c) show the longitudinal ADMR curves with applying magnetic field of 3T, and the ADMR was defined as ADMR=$[ρ−ρ(0\ \text{deg})]/ρ(0\ \text{deg})$ [15]. The ADMR of the two devices show the expected behavior of the SMR, in agreement with the earlier report in Pt/YIG bilayers, and the values in the treated and non-treated devices are about $5.723×10^{-4}$ and $7.814×10^{-4}$, respectively. One can find that the SMR of Pt/YIG is 1.4 times larger than that in Pt/YIG$^+$ device. In order to further investigate the anomalous SMR reduction in Pt/YIG$^+$ device, we carried out the ADMR measurements at the temperature range from 5 to 300 K. The temperature dependent ADMR of Pt/YIG and Pt/YIG$^+$ bilayers in the $β$ scan with 3 T field are shown in Figs. 3 (a) and (b), which all can be well fitted by the SMR mechanism. The Fig. 3(c) displays the temperature dependent SMR of the two devices. It is obvious that the SMR changes non-monotonically with decreasing temperature, which is in debate since it might stem from the competition of two physical mechanisms: the spin Hall effect-induced magnetoresistance (SHE-MR) and the magnetic proximity effect-induced magnetoresistance (MPE-MR) [16]. The MPE-MR becomes evident at relatively lower temperature due to the magnetization induced by the MPE, while at higher temperature, the thermal fluctuations will dominate, disrupting the spontaneous Pt magnetization and eliminating the MPE-MR. More interestingly, the temperature dependence of SMR in Pt/YIG bilayers in the previous reports was weak, while the sharp drop of SMR below 100 K in our Pt/YIG bilayers would be discussed latter via systematic ADMR measurements with varying the thickness of Pt. Furthermore, we have defined the ratio ΔSMR as ΔSMR=[SMR(YIG)-SMR(YIG$^+$)]/SMR(YIG), and the temperature dependence is shown in Fig. 3(d). It seems to be constant below 100 K and linearly increases from 100K to 300K. The anomalous temperature dependence behavior may be related to the magnon excitations, which should also be suppressed by the MPE and increase with temperature. To further verify our speculation about the magnon excitation modulated SMR, we have carried out the Pt thickness (*t*) dependent measurements. The unusual ΔSMR fluctuation curve with increasing *t* was shown in Fig. 4 (a). We can find that the ΔSMR is irrelevant with the bulk spin Hall angle (SHA) and spin diffusion length (SDL) of Pt that should exhibit a peak value at *t* ~ 3 nm [17]. Therefore, the only possibility of the unusual ΔSMR should originate from the change of the interface SMC due to the suppression of magnon transport after etching.

To qualitatively analyze the experimental results, we employ a simulation within the spin drift-diffusion theoretical framework. According to the SMR theory, the longitudinal resistivities of the Pt layer are given by $\rho = \rho_0 + \Delta\rho_0 + \Delta\rho_1(1-m_y^2)$, where $m(m_x,m_y,m_z)=M/M_s$ are the normalized projections of the magnetization of the YIG film to the three main axes, $M_s$ the saturated magnetization of the YIG, $\rho_0$ is the Drude resistivity, $\Delta\rho_0$ accounts for a number of corrections due to the SHE and $\Delta\rho_1$ is the main SMR term. The SMR is quantified by [10,18]:

$$\Delta\rho/\rho \approx \frac{\theta_{SH}^2 \lambda_{sd}}{t} \frac{\tanh^2(t/2\lambda_{sd})}{1/(2\rho\lambda_{sd}G_r)+\coth(t/\lambda_{sd})} \quad (1)$$

Where $\lambda_{sd}$, $\theta_{SH}$, $t$, and $G_r$ are the spin Hall angle, the spin diffusion length, the Pt layer thickness, and the real part of the SMC at the YIG/Pt interface, respectively. The thickness dependence of the longitudinal resistance is shown in Fig. 4(b), and the product of the longitudinal resistivity $\rho$ and the Pt film thickness $t$ is found to change linearly with the film thickness as shown in the inset of Fig. 4(b). Considering the product is found to well obey the equation $\rho t = \rho_b t + \rho_s$ with the bulk resistivity $\rho_b$ and the interfacial resistivity $\rho_s$, which are determined to be 8.0 μΩ.cm and 32.0 μΩ.cm$^2$ based on the fitted lines, respectively. The bulk resistivity is close to the value of 10.0 μΩ.cm of the bulk Pt [19]. Here, we have used the SHA, the SDL and the $G_r$ as 0.07, 1.5 nm and 5×10$^{15}$ Ω$^{-1}$ m$^{-1}$ [16,18,20]. We can find a large discrepancy between the fitted and the measured results as shown in Fig. 4(c). However, the variation trend of the SMR could be fitted with giving SHA = 0.131 and SDL = 0.864 nm as shown in Fig. 4(d), and the large SHA should stem from the interfacial contribution [21]. Therefore, the sharp drop of SMR ratio below 100 K in our Pt/YIG bilayers could be derived from the large interfacial SHA.

Notably, the Pt/YIG$^+$ device should have similar SHA and SDL with Pt/YIG device. We can fit the $G_r$ in Pt/YIG$^+$ bilayers through giving SHA 0.131 and SDL 0.864 nm with Eq. (1). According to the fitted results as shown in Fig. 5(a), we can determine the $G_r$ at the Pt/YIG$^+$ interface is one order of magnitude smaller than Pt/YIG interface. Furthermore, the Hanle magnetoresistance (HMR) cloud modulate the resistance of the HM layer with *H* instead of *M*, exhibiting the similar angular dependent behavior with SMR: no resistance correction is observed for *H* parallel to *σ*, whereas a resistance increase is obtained for *H* perpendicular to *σ* [22,23]. Therefore, the anomalous SMR reduction in our devices may also stem from the different HMR. Notably, the HMR is due to the spin precession around the external magnetic field $H$, leading to the spin relaxation, so we could distinguish SMR and HMR from the field dependent MR measurement. As shown in Fig. 5(b), we can find that the distinction of HMR is negligible as compared with SMR in Pt/YIG and Pt/YIG$^+$ devices at $H$ =3 T. Recently, Y. Dai, *et al.* have found that the simulation method by Eq. (1) exhibits discrepancy because the SHA is fluctuation with varying the thickness of Pt layer. They put forward the electron diffusion coefficient (EDC) and SDL that could be precisely estimated through the ratio of HMR and SMR [21]. Therefore, we would use it to determine the EDC and SDL in the Pt/YIG devices. Then, the SMC at Pt/YIG$^+$ interface could be calculated with the similar EDC and SDL in Pt/YIG. Notably, the HMR/SMR ratio is independent of the SHA and thus the HMR/SMR ratio reads [21]:

$$HMR/SMR = \mathrm{Re}\left\{\frac{1}{\sqrt{1+\frac{i\lambda_{sd}^2 g\mu_B B}{D\hbar}}} \frac{\tanh^2(t\sqrt{\frac{1}{\lambda_{sd}^2}+\frac{ig\mu_B B}{D\hbar}}/2)}{\tanh^2(t/2\lambda_{sd})} \frac{1+2\rho_{xx}\lambda_{sd}G_r \coth(t/\lambda_{sd})}{1+\frac{2\rho_{xx}G_r \coth(t\sqrt{\frac{1}{\lambda_{sd}^2}+\frac{ig\mu_B B}{D\hbar}})}{\sqrt{\frac{1}{\lambda_{sd}^2}+\frac{ig\mu_B B}{D\hbar}}}}\right\} \quad (2)$$

where $g$, $\mu_B$, $D$, and $B$ are the Landé factor, the Bohr magneton, the EDC, and the magnetic induction intensity, respectively. The SDL and the EDC are determined to be 2.02 nm and $4 \times 10^{-6}$ m$^2$s$^{-1}$, respectively, where the longitudinal resistivity $\rho$ = 25.5 μΩ.cm, $G_r$ = $5 \times 10^{15}$ Ω$^{-1}$m$^{-1}$, and the Landé factor $g \approx 2.0$. As shown in Fig. 6, if one assigns the $\lambda_{sd}$ and the $D$ with other values that deviate from 2.02 nm and $4 \times 10^{-6}$ m$^2$s$^{-1}$, the fitted results cannot reproduce for all the samples. Based on the determined $\lambda_{sd}$ and $D$, the $G_r$ for Pt/YIG$^+$ device is five times smaller than Pt/YIG device.

However, there is still a problem which needs to be further clarified. From Fig. 6 (c), we can find that the fitted curves are insensitivity to the SMC which is used as the only free parameter. It is difficult to point out which of the two fitted methods is more precise. Here, we put forward a simple model to further explain this result. According to the report by X. P. Zhang in Ref. [23], the $G_r$ should be read as $G_r = e^2v_F(1/\tau_P-1/\tau_T) = e^2v_F/\tau_P + G_s$, where $e$, $v_F$, $\tau_P$ and $\tau_T$ are the elementary charge ($e > 0$), the density of states per spin species at the Fermi level, the longitudinal and transverse spin relaxation times per unit length for the itinerant electron, respectively. We note that $G_r$ represents the difference between the longitudinal spin relaxation with transverse spin relaxation and it does not have a physical meaning on its own. The $G_s=-e^2v_F/\tau_T$ originates entirely from spin-flip processes and associates with magnon emission and absorption [24,25]. Therefore, magnon transport could affect the $G_r$. In the Pt/YIG$^+$ device, the part of YIG film around the Hall bar is etched, which produces infinite high barriers at two sides of the Hall bar and suppresses the in-plane magnon transport. Obviously, it will increase the magnon accumulation and suppress the spin absorption at Pt/YIG$^+$ interface, which will reduce the $G_r$ and the corresponding SMR [26]. Q. Shao *et al.* also found that the measured SOT efficiency was significantly enhanced with increasing the FMIs (TmIG) thickness, which is completely different from the FMs based devices. Similarly, we also found that the SMR is significantly enhanced with increasing the YIG thickness as shown in Fig. 6(d). Therefore, we also ascribe the modulated SMR to the magnon excitations because the thicker YIG is benefit for magnon diffusion, reducing magnon accumulation [27].

In conclusion, we have found the SMR in Pt/YIG-based devices was dramatically altered when the part of YIG film out of the hall devices was etched by the Ar+-ion milling. At room temperature, the SMR effect in the treated device was smaller than in the non-treated one. According to theoretical simulation, we attributed this phenomenon to the reduction of the $G_{\uparrow\downarrow}$ at the Pt/YIG interface induced by the suppressed magnon transport. Our experimental results pointed out that the SMR and the MMR that were inter-connected, and the SMR could be modulated via magnon excitation/suppression even though the magnitude of MMR from magnon excitations was much smaller than the SMR. Our findings provide a new approach for modulating SMR for spintronic applications.

**Acknowledgements:** This work was partially supported by the National Science Foundation of China (Grant Nos. 51971027, 51731003, 51671019, 51602022, 61674013, 51602025), and the Fundamental Research Funds for the Central Universities (FRF-TP-19-001A3).

**Figure captions**

Figure 1 (a) Sample and measurement configurations with the definition of the axes. (b) The XRD *ω-2θ* scans of the YIG film grown on GGG substrate. (c) High-resolution XRD reciprocal space maps of the YIG film grown on GGG substrate. (d) Field dependence of out-of-plane magnetization (black) and in-plane magnetization (red) of YIG film.

Figure 2 (a) Notations of different rotations of the angular *α*, *β*, and *γ*. (b) and (c) ADMR in Pt/YIG and Pt/YIG$^+$ devices at 300 K with ***H*** = 3 T in the three rotation planes ( *α*, *β* and *γ* ).

Figure 3 (a) and (b) ADMR curves measured in Pt/YIG and Pt/YIG$^+$ devices with varying *β* at different temperature, and the applied magnetic field is 3 T. (c) Temperature dependence of the ADMR of Pt/YIG (black) and Pt/YIG$^+$ (red) devices. (d) Temperature dependence of the ΔSMR.

Figure 4 The ΔSMR (a) and longitudinal resistance in the two devices (b) at room temperature with varying Pt layer thickness (*t*). The inset of (b) shows the Pt layer

thickness dependence of the product of the longitudinal resistivity and thickness in the two devices. (c) and (d) The Pt layer thickness dependence of the experimental and fitted SMR in Pt/YIG with different parameters.

Figure 5 (a) The experimental and fitted SMR in Pt/YIG$^+$ with varying the Pt layer thickness, and $G_r$ is parameter for fitting. (b) The MR versus the external magnetic field ***H*** along *Y* axis and *Z* axis.

Figure 6 (a) and (b) The measured and fitted of HMR/SMR (***H***) curves in Pt/YIG device. In panel (a), the red lines refer to the fitted results where the SDL and the EDC are free parameters. The blue and pink lines refer to the fitted results with fixed but different EDC. In panel (b), the red lines refer to the fitted results where the SDL and the EDC are free parameters. The blue and pink lines refer to the fitted results with fixed but different EDL. (c) The measured and fitted of HMR/SMR (***H***) curves in Pt/YIG$^+$ device. The red lines refer to the fitted results where the SMC is free parameter. The blue and pink lines refer to the fitted results with fixed but different SMC. (d) Temperature dependence of the ADMR of Pt/YIG (20 nm) and Pt/YIG (60 nm) bilayers.

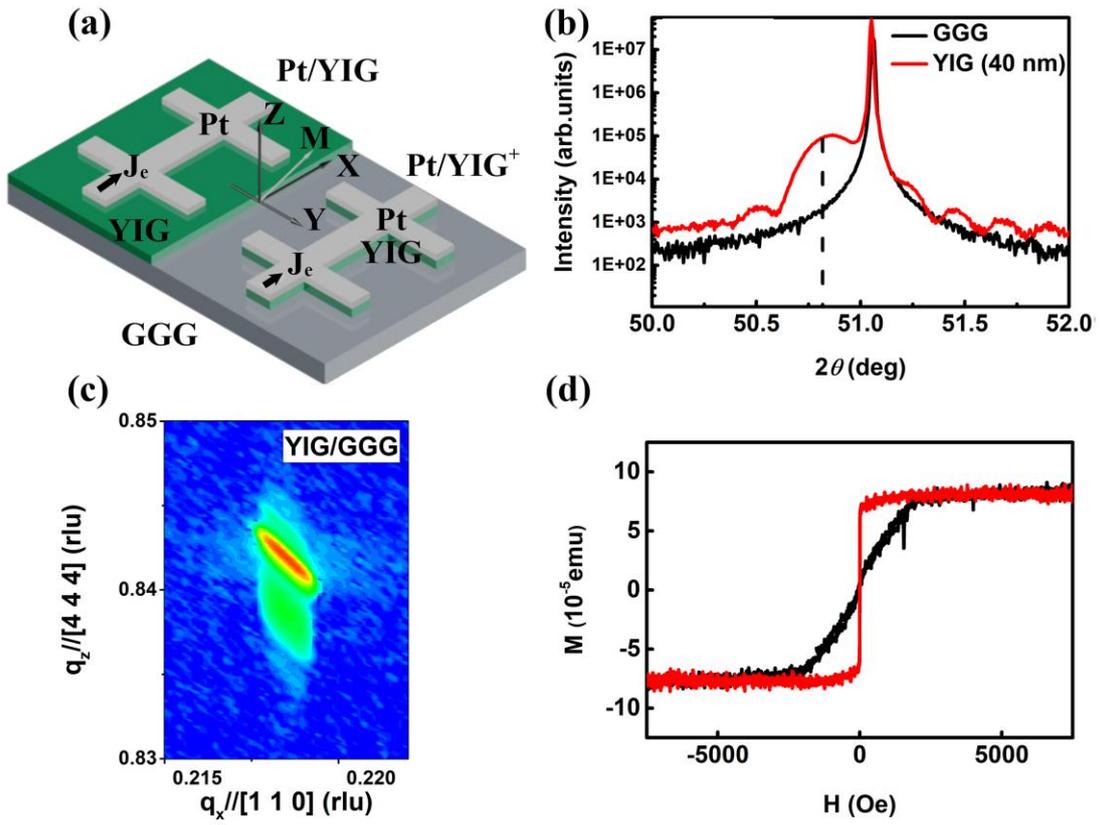

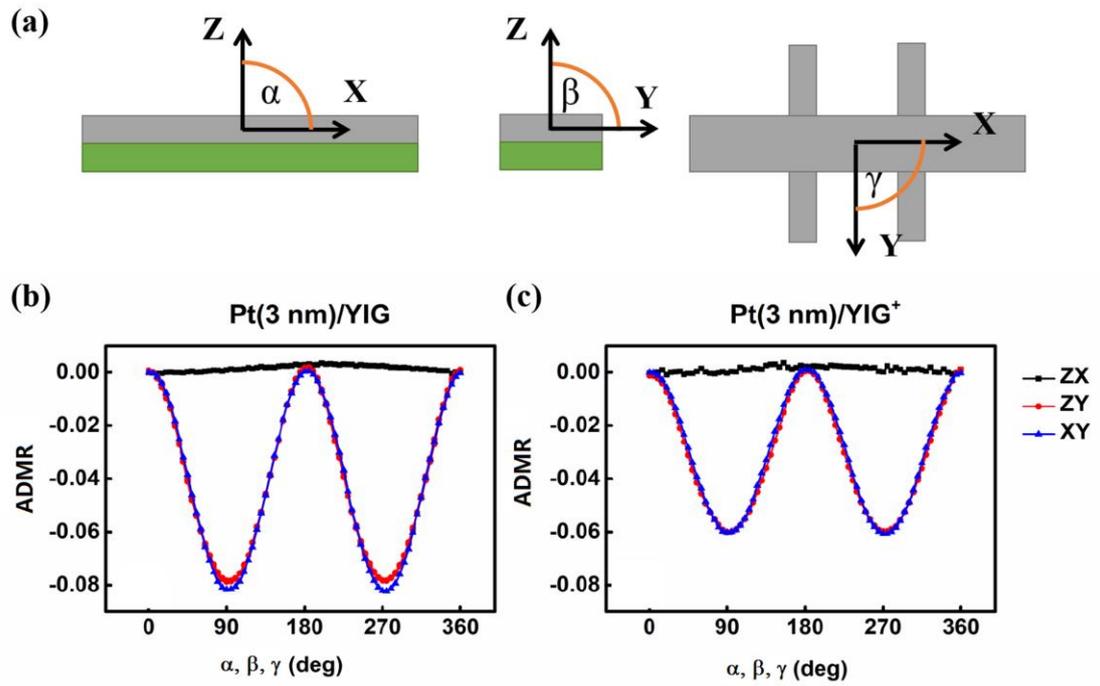

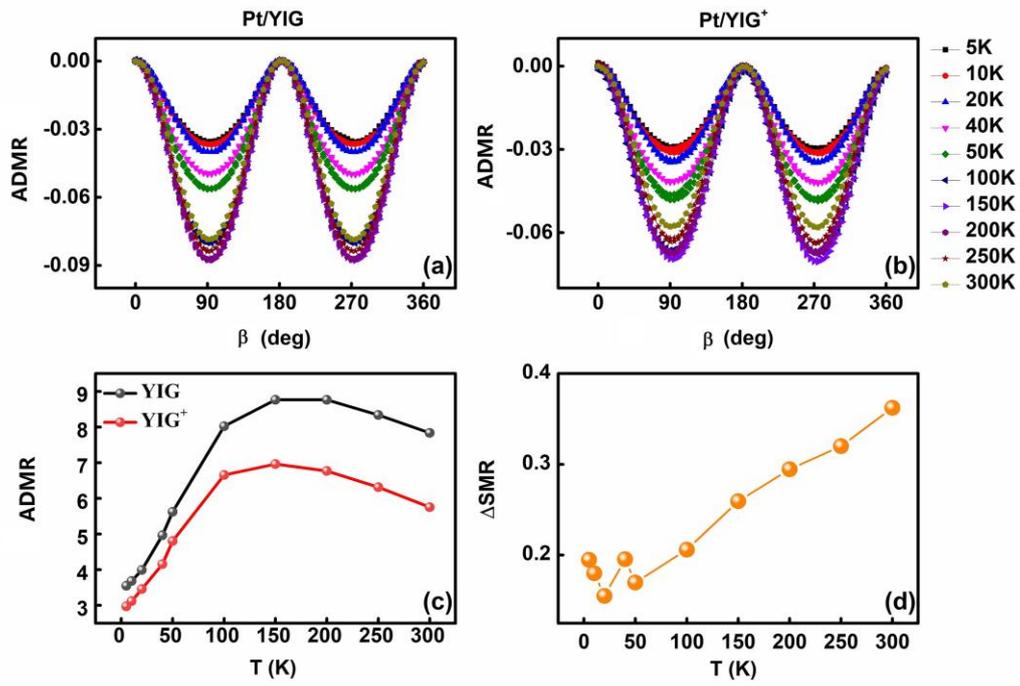

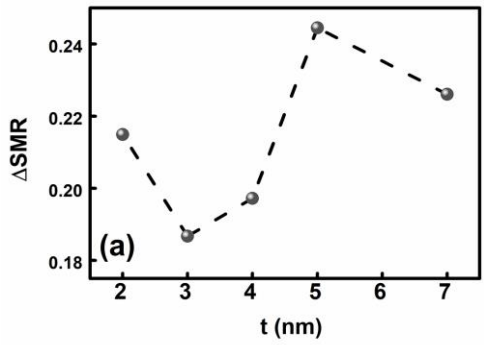
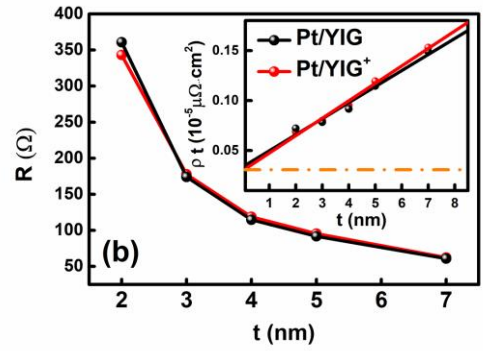
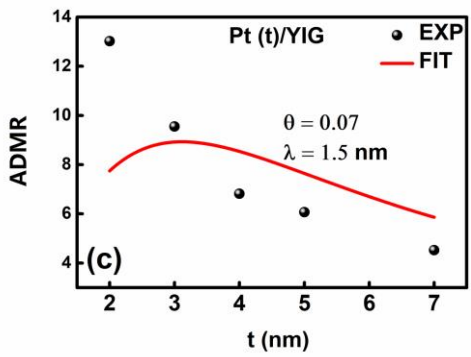
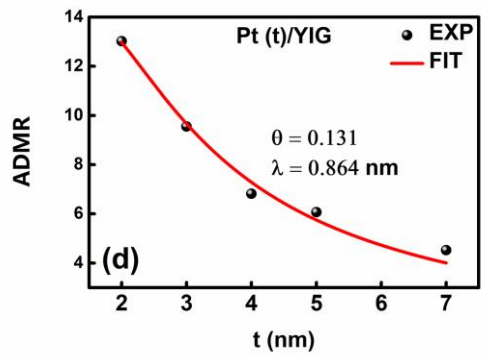

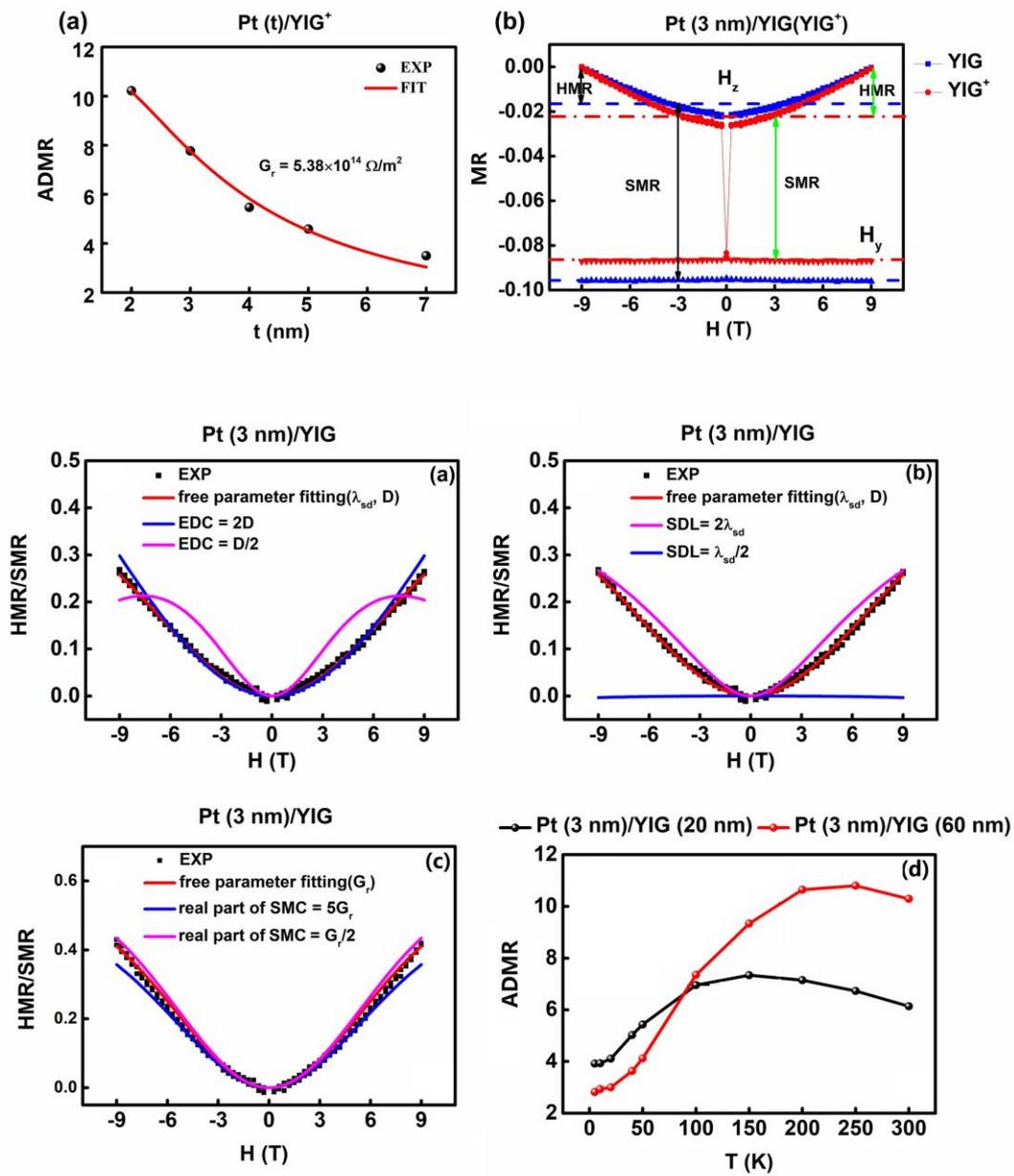